\documentclass[conference,10pt]{IEEEtran}
\usepackage{epsfig,rotating,setspace,latexsym,amsmath,epsf,amssymb,amsfonts,bm,subfigure,epstopdf,cite,authblk,bbm,mathtools,amsthm,algorithm,color,systeme,mathtools,amsmath,derivative,bigints}
\usepackage[utf8]{inputenc}
\usepackage[T1]{fontenc}
\usepackage[thinc]{esdiff}
\usepackage[noend]{algpseudocode}

\theoremstyle{plain}
\newtheorem{theorem}{Theorem}

\newtheorem{corollary}{Corollary}

\newtheorem{assumption}{Assumption} 
 
\newenvironment{Proof}[1]{\medskip\par\noindent{\bf Proof:\,}\,#1}{{\mbox{\,$\blacksquare$}\par}}

\algrenewcommand\algorithmicforall{\textbf{foreach}}
\algrenewcommand\algorithmicindent{.8em}

\IEEEoverridecommandlockouts
\allowdisplaybreaks

\begin{document}

\title{When to Preempt in a Status Update System?}

\author{Subhankar Banerjee \qquad Sennur Ulukus\\
\normalsize Department of Electrical and Computer Engineering\\
\normalsize University of Maryland, College Park, MD 20742\\
\normalsize \emph{sbanerje@umd.edu} \qquad \emph{ulukus@umd.edu} }
	
\maketitle

\begin{abstract}
We consider a time-slotted status update system with an error-free preemptive queue. The goal of the sampler-scheduler pair is to minimize the age of information at the monitor by sampling and transmitting the freshly sampled update packets to the monitor. The sampler-scheduler pair also has a choice to preempt an old update packet from the server and transmit a new update packet to the server. We formulate this problem as a Markov decision process (MDP) and find the optimal sampling policy. We find a sufficient, and also separately a necessary, condition for the always preemption policy to be an optimal policy. We show that it is optimal for the sampler-scheduler pair to sample a new packet immediately upon the reception of an update packet at the monitor. We propose a \emph{double-threshold} sampling policy which we show to be an optimal policy under some assumptions on the queue statistic. 
\end{abstract}

\section{Introduction}
\begin{figure}[t]
	\centerline{\includegraphics[width = 1\columnwidth]{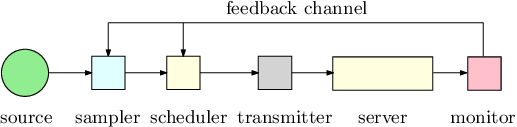}}
	\caption{A sampler-scheduler pair decides the packet flow through the server such that the monitor receives as fresh information as possible about the source. The sampler-scheduler pair decides when to take a fresh sample, and when (if at all) to preempt a packet being served at the queue.}
	\label{fig1}
	\vspace*{-0.4cm}
\end{figure}

We consider a network model which comprises a source, a sampler, a scheduler, a transmitter, a preemptive server and a monitor. The source is a stochastic process which the monitor aims to track in real-time. At a given time slot $t$, the sampler samples an update packet and aims that the monitor receives this update packet so that the monitor has a fresh information about the source. Let us consider that at time $t$, when the sampler samples a new update packet, the server is busy serving a previously sampled update packet. Then, the scheduler has to decide whether to preempt the old update packet and transmit the new update packet, or to keep transmitting the old update packet and discard the new update packet. The preemptive nature of the server considered in this paper gives an extra degrees of freedom to the scheduler to minimize the age of information compared to a non-preemptive server. A sampling algorithm $\pi$ is composed of a sampling policy of the sampler as well as a preemption policy of the scheduler. Our goal is to devise a sampling and preemption algorithm such that the monitor has as fresh information as possible about the source. We characterize this freshness by the well-studied metric of age of information \cite{Kosta17agesurvey, SunSurvey, YatesSurvey}. Fig.~\ref{fig1} provides a pictorial representation of the considered network model.

The sampling problem in the context of age of information minimization has been studied in the literature, see e.g.,~\cite{sun2019sampling, bedewy2021optimal}. These works consider non-preemptive servers, i.e., a packet being served cannot be dropped from the server and the scheduler has to wait until the monitor receives the currently served packet. Due to that, these works can only optimize the initial sampling times. In contrast, in this paper, we can optimize the initial sampling times as well as the preemption times. Now, between two successful receptions of the update packets at the monitor, there can be multiple preemptions. Thus, in principle, in this paper, we have to deal with countably infinite optimization parameters, which makes the studied problem challenging. 

The sampling problem to minimize the age of information with possible preemption has also been studied in the literature under different network settings, see e.g., \cite{arafa2019aller, wang2019preempt, yates2018age2}. Reference \cite{arafa2019aller} considers a network model which is similar to our model, except that it is in continuous time whereas ours is in discrete time. To combat the countably infinite optimization parameters, \cite{arafa2019aller} finds the optimal policy from the set of policies for which the preemption threshold is fixed and independent of the sample indices, i.e., if the service time of a packet in the server exceeds a fixed threshold then the scheduler-sampler pair drops the current packet and samples a fresh packet from the source and submits it to the server. Thus, \cite{arafa2019aller} optimizes two parameters to find the optimal policy, the initial sampling time, i.e., when to sample an update packet when the server is empty, and the preemption threshold. In contrast, we consider the set of all causal policies for the discrete time setting. 

In this paper, first, we formulate a Markov decision process (MDP) for the considered optimization problem. Then, we show that there exists a stationary sampling and preemption policy which is optimal for the considered problem. Next, we find several structural properties for the optimal policy for the MDP. Namely: We find that the zero-wait sampling policy is optimal. We find a necessary, and separately a sufficient, condition for the \emph{always-preempt} policy to be an optimal policy. We prove various threshold structures of an optimal policy. Next, we propose a \emph{double-threshold} based policy, in which, the first threshold determines the continuous preemption duration, and when the continuous preemption period is over, the second threshold determines the waiting period of a packet in the server before preempting it. We show that under some assumptions on the queue statistics, the proposed policy is optimal. 

\section{System Model and Problem Formulation}
We consider a time slotted model, where a sampler can generate an update packet at any time slot it wishes, i.e., we consider a generate-at-will (GAW) model. The sampler aims to minimize the age of a monitor by delivering the fresh update packets. We model the communication channel as an error-free preemptive queue. As we are interested in freshness, at time slot $t$, the sampler samples a new packet only if either the queue is empty or the sampler decides to preempt the existing packet in the queue \cite{sun2019sampling}. We denote a sampling policy as $\pi$. We consider a set of causal policies, i.e., the policies which depend only on the past observations and decisions, denoted by $\Pi$. We denote the generation time of the $i$th update packet corresponding to a sampling policy $\pi$ as $S_{i}^{\pi}$ and the delivery time of the $i$th packet as $D_{i}^{\pi}$. If the $i$th packet is preempted by the server, then we define $D_{i}^{\pi} = \infty$. At time $t$, we define the age of information of the monitor corresponding to a sampling policy, $\pi$ as $v^{\pi}(t) = t-S^{\pi}_{i}$, where $i = \sup{\{j:D_{j}^{\pi}\leq t\}}$. Thus, a sampling policy $\pi$ is completely defined as $\pi = (S_{1}^{\pi}, S_{2}^{\pi}, \cdots)$, and we are interested in solving the following problem,
\begin{align}\label{eq:2}
    \inf_{\pi\in\Pi} &\limsup_{T\rightarrow \infty} \frac{1}{T} \sum_{t=1}^{T} \mathbb{E}_{\pi} [v^{\pi}(t)|v^{\pi}(1)=v],
\end{align}
where $v$ is the initial age of the monitor irrespective of $\pi$.

Note that, it is not always guaranteed that the monitor will receive a certain sampled packet, as the sampler can preempt an update packet from the queue. When the monitor successfully receives an update packet for the $i$th time, we denote that event as $\mathcal{E}_{i}^{\pi}$, $i\in{\{0,1,2,\cdots\}}$. Without loss of generality, we assume that the event $\mathcal{E}_{0}^{\pi}$ occurs at time $t=0$. We denote the number of sampled update packets between the event $\mathcal{E}_{i}^{\pi}$ and $\mathcal{E}_{i+1}^{\pi}$ as $M_{i}^{\pi}$. Note that, $M_{i}^{\pi}=0$ implies that the sampler does not preempt any update packet after the $i$th sample packet received by the monitor until the $(i+1)$th sample packet is received by the monitor. We denote the time when the event $\mathcal{E}_{i}^{\pi}$ occurs with $\bar{D}_{i}^{\pi}$. We assume that $S_{0}^{\pi}=0$, irrespective of any policy $\pi$. The sample index of the update packet, upon the reception of which the event $\mathcal{E}_{i}$ occurs, is
\begin{align}\label{eq:k}
    k= \sum_{j=0}^{i-1} M_{j}^{\pi} + i.
\end{align}

We assume that the time taken to deliver the $k$th sampled packet, defined in (\ref{eq:k}), to the monitor is $Y_{i}$. We call $Y_{i}$ as the service time of the queue. We assume that the service time is independent and identically distributed over the index of the delivered sample $i$. Thus, for notational convenience, we use $Y$ as the service time. We assume that $\mathbb{P}(Y=i)=p_{i}$. In this work we assume that $p_{1}>0$. Note that, if a packet is at the server for $k$ time slots, i.e., if the age of the packet at the server is $k$, then we denote the probability with which the monitor receives that packet with $q_{k+1}$, i.e., $q_{k+1} = \frac{p_{k+1}}{\sum_{i=k+1}^{\infty}p_{i}}$. Note that the delivery time of a sampled packet to the monitor is independent of the policy and only depends on the queue statistics. Thus, the time when the event $\mathcal{E}_{i}$ occurs is $(S_{k}^{\pi} + Y_{i})$. We assume that $\mathbb{E}[Y_{i}] < \infty$. Now, we define the following quantities to solve (\ref{eq:2}). We define the first sampling time after the occurrence of the event $\mathcal{E}_{i}$ as $X_{i}^{\pi}$. Note that, there are $M^{\pi}_{i}$ number of samples between the event $\mathcal{E}_{i}$ and $\mathcal{E}_{i+1}$. Thus, using (\ref{eq:k}),
\begin{align}
    X_{i}^{\pi} = S_{k+1}^{\pi} -  S_{k}^{\pi}.
\end{align}
Moreover, we denote the time interval between the $j$th sample and the $(j+1)$th sample in between the events $\mathcal{E}_{i+1}^{\pi}$ and $\mathcal{E}_{i}^{\pi}$, with $Z_{j,i}^{\pi}$, for $j\in{1,2,\cdots, (M_{i}^{\pi}-1)}$. From \cite{sun2019sampling}, we know that it is not optimal to sample a packet when the queue is busy, thus, we assume that the waiting time for an update packet to get served by the queue is $0$. In other words, the sampler decides to sample an update packet only after a successful transmission of the serving packet in the queue, or if the sampler decides to preempt the serving packet in the queue. Thus, we can represent a scheduling algorithm as,
\begin{align}\label{eq:is5}
\pi=(X_{0}^{\pi}, Z_{1,0}^{\pi}, \cdots, Z_{M_{0}^{\pi}-1,0}^{\pi}, X_{1}^{\pi},\cdots).
\end{align} 
Note that, if a policy $\pi$ never samples an update packet irrespective of the age of the monitor, then the average age of the problem in (\ref{eq:2}) goes to $\infty$ as $T$ goes to $\infty$. Thus, we consider that all the policies in the policy set $\Pi$ sample an update packet at some finite age of the monitor.

Due to the space limitations here, we provide proofs for some selective results; the rest of the proofs will be provided in a journal version, which will be posted on arXiv.

\section{Optimal Scheduling Algorithm}
In this paper, we assume that whenever the scheduler preempts an update packet from the server, the scheduler-sampler pair immediately samples a new update packet from the source and decides to transmit this packet to the monitor. In the next theorem, we show that this is indeed the optimal thing to do. In other words, it is not optimal to wait for some time to sample a new packet when the scheduler preempts an update packet from the server. 
\begin{theorem}
    It is always optimal for the scheduler-sampler pair to sample a new update packet from the source and transmit the packet to the monitor immediately after the scheduler decides to preempt a currently serving update packet.
\end{theorem}
Before formulating the problem in (\ref{eq:2}) in an MDP, we first define the components of the MDP.

\textit{State:} We define a state of the system as a two dimensional vector $s=(v_{1}, v_{2})$ where $v_{1}$ is the age of the monitor and $v_{2}$ is the age of the packet in the server. Similarly, we define the state of the system at time $t$ as $s(t)=(v_{1}(t), v_{2}(t))$. We define the set of all the states $s$ as the state space of the problem $\mathcal{S}$. Note that, $\mathcal{S}$ is a countably infinite set. If the server is empty and the age of the monitor is $v_{1}$, we represent the state of the system with $(v_{1},\infty)$. Note that, the age of the monitor can never be less than the age of the packet in the queue. Thus, we always assume that $s=(v_{1}, v_{2})$, such that $v_{1}\geq v_{2}$. If with probability $1$, the service time $Y$ is finite, in other words, if there exists a finite positive integer $L$, such that the random variable is bounded by $L$, i.e.,
\begin{align}\label{eq:bound_rand}
    L = \inf \Big\{k: \sum_{i=1}^{k} \mathbb{P}(Y\!=\!i) =1\Big\},
\end{align}
then, the age of the packet in the queue can never be greater than $k$. Thus, we modify the state space as
\begin{align}\label{eq:state_space}
    \!\!\mathcal{S} \!=\! \{(v_{1},v_{2})\!\cup\!(v_{1},\infty): v_{1}\geq v_{2}, (v_{1},v_{2})\in{\mathbb{N}^{2}}, v_{2} \leq L \}.
\end{align}

\textit{Action:} At time slot $t$, the sampler-scheduler pair has three actions to choose from. We write the action at time $t$ as $a(t) \in \{0,1,2\}$. Here, $a(t)=0$ means that at time $t$, the server is empty and the sampler does not sample a new update packet. Next, $a(t)=1$ means that at time $t$, the sampler samples a fresh update packet and the scheduler decides to transmit it to the monitor, i.e., if at time slot $t$ the server is empty then the transmitter submits this packet to the server, or if the server is busy serving a staler packet, then the scheduler preempts that old update packet and submits the fresh update packet to the server. Finally, $a(t)=2$ means that at time $t$, the server is serving an update packet and the scheduler decides to keep transmitting that update packet.

\textit{Transition probabilities:} If the system is in state $s$ and the sampler-scheduler pair decides to choose an action $a$, then we denote the probability with which the system goes to state $s'\in{\mathcal{S}}$ as $P_{a}(s,s')$. The transition probabilities depend on the statistics of the service time $Y$.

\textit{Cost:} If the state of the system is $s\in{\mathcal{S}}$, and if the sampler-scheduler takes an action $a\in{\{0,1,2\}}$, then we define the cost incurred by the system as $C(s,a)=v_{1}$.

Thus, (\ref{eq:2}) can be reformulated as,
\begin{align}\label{eq:3} 
    \inf_{\pi\in{\Pi}} & \limsup_{T\rightarrow \infty} \frac{1}{T} \sum_{t=1}^{T} \mathbb{E}_{\pi} [C(s(t),a(t))|s(1)=(v,\infty)].
\end{align}
In (\ref{eq:3}) we assume that the initial age of the system is $v$ which is independent of the policy, and initially the queue is empty. We denote the MDP with the above mentioned state space, action space, cost and transition probabilities, with $\Delta$. 

In the next theorem, we show that there exists a stationary policy which optimally solves the problem in (\ref{eq:3}). 

\begin{theorem}\label{th:1}
    There exists a stationary policy which is optimal for the following problem,
 \begin{align} 
    \inf_{\pi\in{\Pi}}  \limsup_{T\rightarrow \infty} \frac{1}{T} \sum_{t=1}^{T} \mathbb{E}_{\pi} [C(s(t),a(t))|s(1)=(v,\infty)].
\end{align}
\end{theorem}

Next, we introduce the discounted MDP, which is well-known in the literature \cite{puterman2014markov, bertsekas2012dynamic}, to prove the structural properties of an optimal solution for (\ref{eq:3}). For $0<\alpha<1$, we define the $\alpha$-discounted cost for an initial state $s$, corresponding to a policy $\pi$ as,
\begin{align}\label{eq:discounted}
    V_{\alpha}^{\pi}(s) = \sum_{t=1}^{\infty} \alpha^{T} \mathbb{E}_{\pi} [C(s(t),a(t)) | s(0)=s].
\end{align}
We define the optimal discounted cost as, $V_{\alpha}(s) = \inf_{\pi\in{\Pi}} V_{\alpha}^{\pi}$. We define the following quantity which will be useful for future presentation,
\begin{align}\label{eq:is10}
    V_{\alpha}(s;a) = C(s,a) + \alpha \sum_{\tilde{s}\in{\mathcal{S}}} P_{a}(s,\tilde{s}) V_{\alpha}(\tilde{s}).
\end{align}
For all states $s\in{\mathcal{S}}$ and $0<\alpha<1$, we define $V_{\alpha,0}(s)=0$ and for $n\geq 1$,
\begin{align}\label{eq:38}
   \!\! V_{\alpha,n}(s)\! =\! \max_{a\in{\{0,1,2\}}} \{C(s,a) + \alpha \sum_{\tilde{s}\in{\mathcal{S}}} P_{a}(s,\tilde{s}) V_{\alpha,n-1}(\tilde{s})\}.\!
\end{align}
From \cite{puterman2014markov, bertsekas2012dynamic}, we know,
\begin{align}\label{eq:is13}
    \lim_{n\rightarrow\infty} V_{\alpha,n}(s) = V_{\alpha}(s).
\end{align}

Now, to prove the crucial results in this paper, we consider a continuous state space $\bar{\mathcal{S}}$, where a state $s=(v_{1},v_{2})\in{\bar{\mathcal{S}}}$, has the same structure as a state $s'\in{\mathcal{S}}$, however, the age of the monitor, i.e., $v_{1}$, can take values from the set of real numbers. We consider an MDP $\bar{\Delta}$ on the state space $\bar{\mathcal{S}}$. The cost and action spaces of this new MDP are the same as the original MDP. We denote the probability of transition from a state $s$ to a state $s'$, where $s,s'\in{\bar{\mathcal{S}}}$, under an action $a$ for the MDP $\bar{\Delta}$ with $\bar{P}_{a}(s,s')$. Thus,
\begin{align}
    & P_{0}((v_{1},\infty),(v_{1}+1,\infty)) \!=\! 1, P_{1}((v_{1},\infty),(1,\infty)) \!=\!q_{1}, \nonumber\\
    & P_{1}((v_{1},v_{2}),(v_{1}+1,1))\
    \!\!= \!\!(1-q_{1}), P_{1}((v_{1},v_{2}),(1,\infty)) = q_{1}, \nonumber\\ 
    &P_{1}((v_{1},\infty),(v_{1}+1,1)) = (1-q_{1}), \nonumber\\
    & P_{2}((v_{1},v_{2}),(v_{2}+1,\infty)) = q_{v_{2}+1}, \nonumber\\ 
    & P_{2}((v_{1},v_{2}),(v_{1}+1,v_{2}+1))=(1-q_{v_{2}+1}).
\end{align}

We denote the optimal discounted cost on the state space $\bar{\mathcal{S}}$, with $\bar{V}_{\alpha}(s)$, $s\in{\bar{\mathcal{S}}}$. Similarly, we denote the continuous state space counterparts of $V_{\alpha}(s;a)$ and $V_{\alpha,n}(s)$, in (\ref{eq:is10}) and (\ref{eq:38}), with $\bar{V}_{\alpha}(s;a)$ and $\bar{V}_{\alpha,n}(s)$, respectively. Due to the structure of the transition probabilities, for $s\in{\mathcal{S}}$ the following holds,
\begin{align}
    \bar{V}_{\alpha}(s) = V_{\alpha}(s).
\end{align}

Next, we show that $\bar{V}_{\alpha}(s)$ is a concave function of $v_{1}$, which will play an important role for the upcoming theorems. 

\begin{theorem}\label{th:4}
    The $\alpha$ discounted value function $\bar{V}_{\alpha}((v_{1},v_{2}))$ is a concave function of $v_{1}$. Similarly, $\bar{V}_{\alpha}(s;a)$ and $\bar{V}_{\alpha,n}(s)$ are also concave functions of $v_{1}$.
\end{theorem}
\begin{Proof}
    We prove this theorem by induction. For $n=1$,
    \begin{align}\label{eq:39}
    \bar{V}_{\alpha,1}(s;a) =\begin{cases}
    v_{1}, & i =\{0,2\}, \\ 
    v_{1} + \lambda, &i=1.
    \end{cases}
    \end{align}
    From (\ref{eq:39}), we see that $\bar{V}_{\alpha,1}(s;a)$ is a linear function of $v_{1}$, $a\in{\{0,1,2\}}$. Thus, $\bar{V}_{\alpha,1}(s) = \min_{a\in{\{0,1,2\}}}\{\bar{V}_{\alpha,1}(s;a)\}$ is a concave function of $v_{1}$. Now, assume that for the induction stage $(n-1)$, the value function $\bar{V}_{\alpha,n-1}(s)$ is a concave function of $v_{1}$. From (\ref{eq:38}),
    \begin{align}
        \bar{V}_{\alpha,n}(s;a) = \begin{cases} v_{1} + \alpha \mathbb{E}[ \bar{V}_{\alpha,n-1}(s')], \quad  i= \{0,2\}, \\ v_{1}+ \alpha\mathbb{E}[ \bar{V}_{\alpha,n-1}(s')], \quad i=1.
        \end{cases}
    \end{align}
    From the induction stage, we say that $\bar{V}_{\alpha,n}(s;a)$ is a concave function of $v_{1}$, $a\in{\{0,1,2\}}$. Thus, $\bar{V}_{\alpha,n}(s) = \min_{\{0,1,2\}} \bar{V}_{\alpha,n}(s;a)$ is a concave function of $v_{1}$. Now, we can take the limit on $n$ and get the desired result.
\end{Proof}

Next, we show that, if the server is empty, then the sampler should sample an update packet immediately and the scheduler should immediately transmit this packet to the monitor. 

\begin{theorem}\label{th:7}
    If the queue is empty then the sampler-scheduler pair always chooses action $a=1$ over action $a=0$.
\end{theorem}

Next, in Theorem~\ref{th:preemp}, we determine a sufficient condition for which the \emph{always-preempt} policy is optimal. 

\begin{theorem}\label{th:preemp}
    If $q_{1}\geq q_{j}$, $j\in{\{2,3,\cdots\}}$, then the always-preempt policy is an optimal sampling policy.
\end{theorem}

Based on Theorem~\ref{th:preemp}, we obtain the next corollary regarding the optimal sampling policy when the channel statistics follow a geometric distribution.

\begin{corollary}
    If the channel statistic follows a geometric distribution, then the always-preempt policy is optimal.
\end{corollary}

Note that, the geometric distribution has the memoryless property, and it also satisfies the condition of  Theorem~\ref{th:preemp}. However, the condition given in Theorem~\ref{th:preemp} is more general than the memoryless property, as we can find queue statistics which are not geometric and satisfy the condition of Theorem~\ref{th:preemp}.

Next, we propose a sampling policy and study its optimality for certain queue statistics. Consider a sampling and transmission policy $\bar{\pi}$ defined as follows: 
\begin{itemize}
\item After every successful transmission of an update packet, the sampler immediately samples a new update packet, and feeds it to the queue. 
\item After one time slot, if this new update packet does not get delivered to the monitor, the server preempts this packet and samples another new update packet. This continuous stretch of immediate preemption process continues till the age of the monitor \emph{exceeds} a certain threshold $v_{th,1}^{\bar{\pi}}$. 
\item After the age of the monitor has exceeded this threshold, the server does not preempt the most recent update packet immediately and keeps transmitting it till the age of the \emph{most recent packet} \emph{reaches} a certain threshold $v_{th,2}^{\bar{\pi}}$. After this time, the most recent packet is dropped by the server, and the sampler samples a new update packet and keeps transmitting it till the age of that packet again reaches threshold $v_{th,2}^{\bar{\pi}}$. This same process continues till the monitor receives an update packet. 
\end{itemize}

\begin{figure}[t]
	\centerline{\includegraphics[width = 0.9\columnwidth]{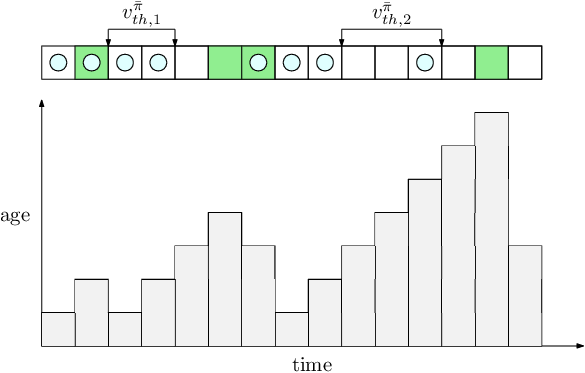}}
	\caption{A pictorial representation of policy $\bar{\pi}$ with ${v}_{th,1}^{\bar{\pi}}=2$, ${v}_{th,2}^{\bar{\pi}}=3$. The green boxes represent the successful transmissions. The cyan circles represent a packet generation. Note that, if a packet is in the server and a new packet is generated by the sampler, then the old packet is preempted from the server, for example, see time slot $4$. Gray curves shows the age at the monitor.}  
	\label{fig2}
	\vspace*{-0.4cm}
\end{figure}

A pictorial representation of this policy is given in  Fig.~\ref{fig2}. We denote the set of all such policies with $\bar{\Pi}$. In the next theorem, we study a sufficient condition for the existence of an optimal policy in the set $\bar{\Pi}$.

\begin{theorem}\label{th:nopreemp}
    If $q_{1}\geq q_{j}$, $j\in{3,4,\cdots}$, then there exists a policy in the policy set $\bar{\Pi}$ which is an optimal sampling policy for the MDP $\Delta$.
\end{theorem}

Note that, different from Theorem~\ref{th:preemp},  Theorem~\ref{th:nopreemp} compares $q_1$ only with $q_3$, $q_4$, $\ldots$, skipping $q_2$. Nonetheless, the conditions for both Theorem~\ref{th:preemp} and Theorem~\ref{th:nopreemp} are not satisfied when the queue statistics is bounded, i.e., $L$ in (\ref{eq:bound_rand}) is finite, as in this case $q_{L}=1$, and it cannot be smaller than $q_1$. In the rest of this paper, we consider bounded $L$, i.e., each packet eventually gets delivered in a finite number of time slots, which is common in practical systems.

In the next theorem, we show that when $v_{2}=(L-1)$, there exists a threshold policy which is an optimal sampling policy. 

\begin{theorem}
    For any state $(v_{1},L-1)$, if an optimal action is $a=2$, then for any state $(v_{1}+x,L-1)$, the action $a=2$ is an optimal action, $x>0$.
\end{theorem}

In the rest of the paper, we assume that the age of the system is upper bounded by a large integer $K$. We choose $K$ large enough such that $K\gg L$. Now,we formally define the bounded state space $\mathcal{S}_{K}$,
\begin{align}\label{eq:state_space_bouded}
  \mathcal{S}_{K} = \Big\{ &(v_{1},v_{2})\cup(v_{1},\infty): v_{1}\geq v_{2}, \nonumber\\
  & (v_{1},v_{2})\in{\mathbb{N}^{2}}, v_{2} \leq L, v_{1}\leq K \Big\}.
\end{align}
Similarly, we define the transition probability from a state $s\in{\mathcal{S}_{K}}$ to a state $s'\in{\mathcal{S}_{K}}$, under the action $a\in{\{1,2\}}$ with,
\begin{align}\label{eq:mobi1473}
    P_{a}(s,s';K) = \begin{cases}
        P_{a}(s,s'), & v_{1}\leq K, v_{1}'\neq v_{1} \\ \sum_{s''\in{\mathcal{S}\backslash\mathcal{S}_{K}}} P_{a}(s,s'') , & v_{1} = K, v_{1}' = v_{1}, \\& v_{2}'=(a-1)v_{2}+1, \\0 , & \textrm{otherwise}.
    \end{cases}  
\end{align}
We denote the MDP with cost of the MDP $\Delta$, transition probability of (\ref{eq:mobi1473}) and state space of (\ref{eq:state_space_bouded}), with MDP $\Delta_{K}$. Next, we introduce the relative value function and relative value iteration for the MDP $\Delta_{K}$. We define $V_{0}(s) = 0$ and $h_{0}(s)=0$, for all $s\in{\mathcal{S}_{K}}$. Consider the following iteration,
\begin{align}\label{eq:mobivernew80}
    V_{n}(s) & =\! \! \min_{a\in{\{1,2\}}}\left\{C(s,a) \!+\!\! \sum_{s'\in{\mathcal{S}_{K}}}\!P_{a}(s,s';K) h_{n-1}(s')\right\} \! \\ 
    h_{n}(s) & =  V_{n}(s) - V_{n}((1,\infty)).
\end{align}
According to \cite[Thm.~4.3.2]{bertsekas2012dynamic}, we have that the sequences $\{V_{n}(s)\}_{n=1}^{\infty}$ and $\{h_{n}(s)\}_{n=1}^{\infty}$ converge to $h^{*}(s)$ and $V^{*}(s)$.

Now, similar to the $\alpha$-optimal value function, we define the relative value function in a continuous state space $\bar{\mathcal{S}}_{K}$, where a state $s=(v_{1},v_{2})\in{\bar{\mathcal{S}}_{K}}$, has the same structure as a state $s'\in{\mathcal{S}_{K}}$, however, the age of the monitor, i.e., $v_{1}$, can take values from the set of real numbers. We define a MDP $\bar{\Delta}_{K}$ on the state space $\bar{\mathcal{S}}_{K}$, with the cost and the action space similar to the MDP $\mathcal{S}_{K}$. We define the transition probability from a state $s$ to $s'$, where $s,s'\in{\bar{\mathcal{S}}_{K}}$, under an action $a$ with $\bar{P}_{a}(s,s';K)$, which has the same structure as $P_{a}(s,s';K)$. We denote the continuous state space counterparts of $V^{*}(s)$, $V^{*}(s;a)$, $V_{n}(s)$, $V_{n}(s;a)$, with $\bar{V}^{*}(s)$, $\bar{V}^{*}(s;a)$, $\bar{V}_{n}(s)$ and $\bar{V}_{n}(s;a)$, respectively. Note that, all the theorems and lemmas regarding the properties of the $\alpha$-optimal value functions hold true for the relative value functions of the MDP $\bar{\Delta}_{K}$. 
 
Next, under two different sets of assumptions, we show that the optimal policy has a threshold structure on $v_{1}$.

\begin{assumption}
   For all $n\in{\mathbb{N}}$, if an  optimal action is $a=2$ corresponding to the value function $\bar{V}_{n}((v_{1},v_{2}))$, then for a state $(v_{1},v_{2}')$ the action $a=2$ is also an optimal action corresponding to the value function $\bar{V}_{n}((v_{1},v_{2}'))$, where $v_{2}'\geq v_{2}$. We also assume that the queue statistics follow the following relation, $q_{1}\leq q_{2}\leq\cdots q_{L-1}$. For example, the queue statistic $[p_1, ~ p_2, ~ p_3]=[0.4, ~ 0.3, ~ 0.3]$, satisfies the condition of Assumption~1, as in this case $[q_1, ~ q_2, ~ q_3]=[0.4, ~ 0.5, ~ 1.0]$.
\end{assumption}

\begin{assumption}
    For a fixed $n$, if the first crossing between the functions $\bar{V}_{n}((v_{1},v_{2});1)$  and $\bar{V}_{n}((v_{1},v_{2});2)$ occurs at a state $(v_{2}+x_{1},v_{2})$, then either the first crossing between the functions $\bar{V}_{n-1}((v_{1},1);1)$ and $\bar{V}_{n-1}((v_{1},1);2)$ occurs at a state $(1+x_{2}, 1)$, or there is no crossing between the aforementioned functions; and either the first crossing between the functions $\bar{V}_{n-1}((v_{1},v_{2}+1);1)$ and $\bar{V}_{n-1}((v_{1},v_{2}+1);2)$ occurs at a state $(v_{2}+x_{3}+1,v_{2}+1)$, or there is no crossing between the aforementioned functions, where $0\leq x_{2}\leq x_{1}+v_{2}$ and $0\leq x_{3} \leq x_{1} $. Similarly, if there is no crossing between the functions $\bar{V}_{n}((v_{1},v_{2});1)$ and $\bar{V}_{n}((v_{1},v_{2});2)$, then there should not be any crossing between the functions $\bar{V}_{n-1}((v_{1},1);1)$ and $\bar{V}_{n-1}((v_{1},1);2)$ and the functions $\bar{V}_{n-1}(v_{1},v_{2}+1;1)$ and $\bar{V}_{n-1}(v_{1},v_{2}+1;2)$. For example, the queue statistics $[p_1, ~ p_2, ~ p_3]=[0.2, ~ 0.3, ~ 0.5]$, satisfies the condition of Assumption~2. In this case, $[q_1, ~ q_2, ~ q_3]=[0.2, ~ 0.6, ~ 1.0]$.
\end{assumption}

\begin{theorem}\label{th:mobi10}
    Under Assumption~1, as well as under Assumption~2, there exists a threshold policy on the age of the monitor which is optimal for the MDP $\bar{\Delta}_{K}$. Specifically, for a state $s=(\bar{v}_{1},v_{2})$ if the action $a=2$ is an optimal action, then for any state $s'=(\bar{v}_{1}+x,v_{2})$, $x>0$, the action $a=2$ is also an optimal action.
\end{theorem}

\begin{theorem}
    If a queue statistics satisfy Assumption~1 or Assumption~2, then there exists a policy in $\bar{\Pi}$, which is optimal for the MDP $\Delta_{K}$.
\end{theorem}

Next, we find the necessary condition for the always-preempt policy to be an optimal policy, when the relative value iteration algorithm is employed. Specifically, for any $(v_{1},1)\in{\mathcal{S}_{K}}$, we find the necessary condition for the following relation to hold,
\begin{align}\label{eq:new135}
    \bar{V}^{*}((v_{1},1);1) \leq \bar{V}^{*}((v_{1},1);2).
\end{align}
We state this result in Theorem~\ref{th:nescond}. For this theorem, let us define $f_{L-1}=1$, $f_{1}=\frac{1}{q_{1}}$ and $f_{i}=0$, $1<i<(L-1)$, for $L\geq 3$.

\begin{theorem}\label{th:nescond}
Consider the following iteration for $L\geq 3$ and for all $1<i<(L-1)$, 
\begin{align}
f_{i} = \min \Big\{1+(1-q_{i+1})f_{i+1},f_{1}\Big\}, 
\end{align}
Then, a necessary condition for the always-preempt policy to be optimal is: $1+ (1-q_{2})f_{2} \geq \frac{1}{q_{1}}$. If $L=2$, then the always-preempt policy cannot be an optimal sampling policy.    
\end{theorem}

\section{Numerical Results}

We call the overall optimal policy for ${\Delta}_{K}$, $\pi^{*}$, which we obtain by performing the relative value iteration method given in (\ref{eq:mobivernew80}). We call the optimal policy in the set $\bar{\Pi}$ obtained by choosing the double-thresholds $v_{th,1}^{\bar{\pi}}$ and $v_{th,2}^{\bar{\pi}}$ optimally, $\bar{\pi}^*$. In the context of our paper, reference \cite{arafa2019aller} considers a policy set with $v_{1,th}^{\bar{\pi}}=1$ and varies $v_{2,th}^{\bar{\pi}}$. We denote the optimal policy from this set of policies as $\tilde{\pi}^{*}$. It is evident that the policy $\bar{\pi}^{*}$ gives at least the same or better performance than $\tilde{\pi}^{*}$. In this section, we compare the average age of the proposed policy $\bar{\pi}^{*}$, the overall optimal policy $\pi^{*}$ and the policy proposed in \cite{arafa2019aller} $\tilde{\pi}^{*}$ for different channel statistics. We have the following:

\begin{itemize}
    \item For the service time statistics $[0.4, ~ 0.2, ~ 0.2, ~ 0.2]$, the average age corresponding to $\pi^{*}$ is $2.4952$, the average age corresponding to $\bar{\pi}^{*}$ is $2.4952$, and the average age corresponding to $\tilde{\pi}^{*}$ is $2.5$.
    \item For the service time statistics $[0.7, ~ 0.1, ~ 0.2]$, the average age corresponding to $\pi^{*}$ is $1.4286$, the average age corresponding to $\bar{\pi}^{*}$ is $1.4286$, and the average age corresponding to $\tilde{\pi}^{*}$ is $1.4286$.
    \item For the service time statistics $[0.05, ~ 0.5, ~ 0.1, ~ 0.3, ~ 0.05]$, the average age corresponding to $\pi^{*}$ is $ 3.8049$, the average age corresponding to $\bar{\pi}^{*}$ is $3.9026$, and the average age corresponding to $\tilde{\pi}^{*}$ is $3.9071$.
    \item For the service time statistics $[0.3, ~ 0.25, ~ 0.1, ~ 0.3, ~ 0.05]$, the average age corresponding to $\pi^{*}$ is $3.2170$, the average age corresponding to $\bar{\pi}^{*}$ is $3.2170$, and the average age corresponding to $\tilde{\pi}^{*}$ is $3.333$.
\end{itemize}
We observe that, except for the third service time statistics, $\bar{\pi}^{*}$ is an optimal policy. We also observe that, except for the second service time statistics, the optimal policy does not lie in the policy set considered by \cite{arafa2019aller}. 

Next, we check the validity of Theorem~\ref{th:nescond}. Consider the following queue statistics, $[0.5, ~ 0.125, ~ 0.125, ~ 0.125, ~ 0.125]$. For this queue, the always-preempt policy is an optimal policy, and it satisfies the condition of Theorem~\ref{th:nescond}.   Now, consider the following queue statistics, $[0.3, ~ 0.175, ~ 0.175, ~ 0.175, ~ 0.175]$. The always-preempt policy is not an optimal policy, and it does not satisfy the condition of Theorem~\ref{th:nescond}. 

\newpage

\bibliographystyle{unsrt}
\bibliography{references}

\end{document}